\documentclass[5p,twocolumn,times,number]{elsarticle}

\usepackage{graphicx}
\usepackage{amsmath}   

\begin{document}

\begin{frontmatter}

\title{Data Quality Monitoring of the CMS Silicon Strip Tracker Detector}

\author[add1]{L.~Benucci\corref{cor}, on behalf of CMS Silicon Strip Tracker Collaboration}
\ead{leonardo.benucci@cern.ch}

\cortext[cor]{Corresponding author}

\address[add1]{University of Antwerp, Belgium}

\begin{abstract}

The Physics and Data Quality Monitoring (DQM) framework aims at providing a homogeneous monitoring environment across various applications related to data taking at the CMS experiment. In this contribution, the DQM system for the Silicon Strip Tracker will be introduced. The set of elements to assess the status of detector will be mentioned, along with the way to identify problems and trace them to specific tracker elements. Monitoring tools, user interfaces and automated software will be briefly described.
The system was used during extensive cosmic data taking of CMS in Autumn 2008, where it demonstrated to have a flexible and robust implementation and has been essential to improve the understanding of the detector. CMS collaboration believes that this tool is now mature to face the forthcoming data-taking era.

\end{abstract}


\end{frontmatter}

The Data Quality Monitoring (DQM) system of an experiment ensures optimal working of a detector and certifies the quality of the data for the physics analysis.

The Silicon Strip Tracker is the largest silicon device ever built and is partioned in three main subparts,
Tracker Inner Barrel and Disks (TIB/TID), Tracker Outer Barrel (TOB) and Tracker End Caps (TEC), resulting in 15,148 strip modules, which correspond to 200\,m$^2$ of sensors surface and 9.3 million readout channels. The DQM system for such a complex, high granularity detector must ensure that any possible problem is identified efficiently at a very early stage of the data acquisition process, so that proper actions can be taken quickly.

In this paper I shall outline the basic implementation of the DQM system for the CMS Silicon Strip Tracker and provide a brief sketch of its operation.

\section{The Silicon Strip Tracker DQM structure}

The DQM system for the Silicon Strip Tracker of the CMS detector has been developed within the CMS Software framework~[1]. Monitoring of detector occurs through a series of histograms (Monitoring Elements, ME) which are filled by accessing information from different levels of the reconstruction. 
This tool is intended to be used both online (during data-taking) for prompt detection of problems in hardware of software, as well as offline (during reconstruction stage) to provide fast feedback on the quality of the data, before the completion of the detailed analysis. 

\begin{figure}
\centering
\includegraphics[width=0.7\linewidth]{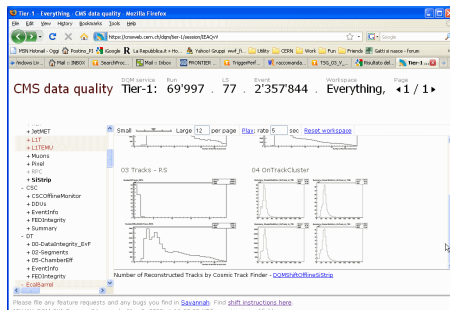}
\caption{A screenshot from a typical view in DQM Graphical User Interface (GUI). Here the distributions for number of tracks and number of recorded hits per track (with two different reconstruction algorithms) are displayed.}
\label{fig:ShiftView}
\end{figure}

\begin{figure}[!Hbtp]
\centering
\includegraphics[width=0.7\linewidth]{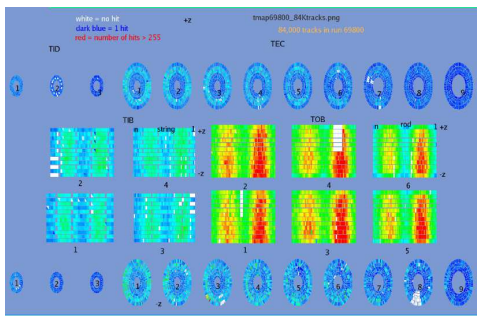}
\caption{`Tracker Map' displaying hit occupancy. Both barrel (rectangular-shaped) and endcap (disk-shaped) subdetectors are represented. Color legenda indicates mean hit occupancy per module. White spots correspond to disconnected modules.}
\label{fig:TkMap}
\end{figure}

The monitoring process is implemented in three steps:

\begin{description}
	\item[DQM producers (source).] Producers retrieve the data from the CMS data acquisition system and ME are booked and then filled with relevant event information. The set of tracker-related variables to be controlled are: ``raw'' data (readout and unpacking errors), ``digis'' and cluster (related or not to a track), track parameters, hit residuals. Due to the large number of silicon modules, about 300,000 histograms are defined for the tracker. They are organized in hierarchical tree-like folder structure (that reflects the tracker geometry) and classified into different categories corresponding to the levels of data processing and reconstruction;
	\item[DQM consumer (client).] It accesses the MEs and performs further analysis. The information from single modules of a subdetector is merged in `Summary plots' and the result can be subjected to quality test, that checks the range of parameters or compare it to reference histogram. The output from both online and offline DQM is eventually archived in ROOT files~[2];	
	\item[Graphical User Interface (GUI).] It provides tools for visualization of the monitoring elements as histograms, detector synoptic views, trend plots etc. Reference distributions can be superimposed and quality test response displayed. The GUI is entirely web-based and can be easily accessed by any user everywhere. For each run a summary as illustrated in Fig.~\ref{fig:SummaryMap} is given, that shows the fraction of active modules per substructure (layer, disk) of a subdetector with a float and a color. In layouts like Fig.~\ref{fig:ShiftView}, a collection of histograms with plots on a layer-basis is exemplified.

 \begin{figure}[!Hbtp]
\centering
\includegraphics[width=0.7\linewidth]{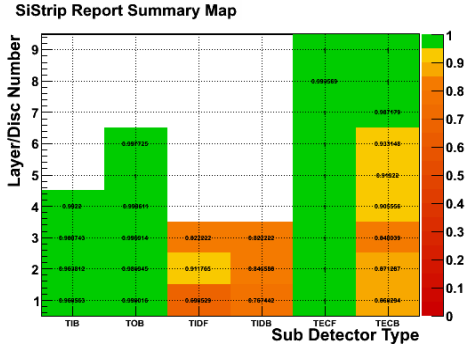}
\caption{Summary report for the fraction of working modules in a run. Horizontal axis has the different tracker subdetectors, vertical axis is for subdetector substructure (layer, disk) and color corresponds to more than 95\% (green), between 85\% and 95\% (dark orange) and below 85\% (pale orange) active modules.}
\label{fig:SummaryMap}
\end{figure}

A two dimensional synoptic view of the tracker, called TrackerMap is shown in Fig.~\ref{fig:TkMap}. It allow identification of several tracker problems easily. In order to quickly identify the position of silicon modules, all the the different subdetectors are represented together. Here the mean occupancy of each module is indicated by a color: with such a kind of tools, ``hot'' (likely noisy) and ``cold'' (likely dead or off) modules can be checked at one glance.

\end{description}

The quality-checking occurs by firstly looking at the summary plot for digis, clusters and tracks. In case a pathology is found, the operator can use the GUI to explore details of the problematic subdetector part (down to the single module level for the offline DQM).

An extensive description of the general DQM system shared by all CMS subdetectors is in Ref.~[3].

\section{DQM operation}

Data monitoring of CMS tracker is organized in online and offline operation steps, followed by a certification procedure. 

\begin{description}
	
 \item[Online DQM.] The DQM system is used online during data taking and accesses a fraction of acquired data. During operation non-expert users (shifters) monitor the data quality of the tracker looking to a set of pre-defined histograms: this guarantees a very effective feedback for tracker problems (mostly hardware-related) and allows to promptly trigger alerts for the experts to take immediate action;
  
	\item[Offline DQM.] In the offline re-processing, all the available statistic can be exploited and a first calibration is used, allowing a detailed data quality analysis. The reprocessing has a latency of some hours and certifies if a run was good. Through an extensive check of a larger set of quantities, shifters are required to re-assess the tracker status of a run and possibly spot reconstruction, calibration or other unexpected problems;
	
	\item[Certification procedure.] Exploiting manual checks from online and offline operators, along with results of quality tests performed by automatic procedures, a set of flags are assigned to the run. They allow to keep track of any new or temporary tracker problem and classify each run according to hardware, reconstruction and calibration conditions. In this way, any user of tracker data can consult the certification results and select suitable runs for specific commissioning or physics analysis tasks. 

\end{description}

During October and November 2008 about 350 million events have been taken with the CMS detector in the cosmic
run at full magnetic field (CRAFT), resulting in about 8 million tracks recorded by the tracker. The DQM system was deployed fully in online and offline environments. With this tool, tracker group could efficiently identify detector 
and reconstruction problems which were fixed accordingly.

\section{Conclusions}

During the last cosmic data taking, DQM system demonstrated to have a flexible and robust implementation and has been essential to improve the understanding of the detector. It was possible to set up a detailed procedure for DQM online and offline operations and test the first prototype of data certification procedure. We believe that this reliable, robust and friendly tool is ready to face the awaited beam collisions stage.

\end{document}